\documentclass{article}\usepackage{natbib,emulateapj}
\newcommand{\beq}{\begin{equation}}
\newcommand{\eeq}{\end{equation}}
\newcommand{\bea}{\begin{eqnarray}}
\newcommand{\eea}{\end{eqnarray}}
\newcommand{\bfig}{\begin{minipage}{3.3in}\bigskip}
\newcommand{\efig}{\bigskip\end{minipage}}
\newcommand{\rem}[1]{ }
\bibpunct[,]{(}{)}{;}{a}{}{,}
\begin{document}

\title{Theory of ``Jitter'' Radiation from Small-Scale Random Magnetic Fields 
and Prompt Emission from Gamma-Ray Burst Shocks}

\author{Mikhail V. Medvedev{}\altaffilmark{1} }

\affil{Harvard-Smithsonian Center for Astrophysics, 60 Garden Street, 
Cambridge, MA 02138}
\altaffiltext{1}{Also at the Institute for Nuclear Fusion, RRC ``Kurchatov
Institute'', Moscow 123182, Russia;
mmedvedev@cfa.harvard.edu; 
http://cfa-www.harvard.edu/\~{ }mmedvede/ }
 
\begin{abstract}
We demonstrate that the radiation emitted by ultrarelativistic electrons 
in highly nonuniform, small-scale magnetic fields is different from synchrotron 
radiation if the electron's transverse deflections in these fields are much 
smaller than the beaming angle. A quantitative analytical theory 
of this radiation, which we refer to as jitter radiation, is developed.
It is shown that the emergent spectrum is determined by statistical properties 
of the magnetic field. The jitter radiation theory is then applied to 
internal shocks of Gamma-Ray Bursts (GRBs). The model of a magnetic field
in GRBs (Medvedev \& Loeb 1999, \apj, 526, 697) is used. 
The spectral power distribution of radiation 
produced by the power-law distributed electrons with a low-energy cutoff
is well described by a sharply broken power-law: $P(\omega)\propto\omega^1$
for $\omega\lesssim\omega_{jm}$ and $P(\omega)\propto\omega^{-(p-1)/2}$
for $\omega\gtrsim\omega_{jm}$, where $p$ is the electron power-law index and 
$\omega_{jm}$ is the jitter break frequency which is independent of the 
field strength but depends on the electron density in the ejecta, 
$\omega_{jm}\propto\sqrt{n}$, as well as on the shock energetics and kinematics.
The total emitted power of jitter radiation is, however, equal to that of 
synchrotron radiation.

Since large-scale fields may also be present in the ejecta, we construct
a two-component, jitter+synchrotron spectral model of the prompt $\gamma$-ray
emission. Quite surprisingly, this model seems to be readily capable of
explaining several properties of time-resolved spectra of some GRBs, such 
as (i) the violation of the constraint on the low-energy spectral index called
the synchrotron ``line of death'', (ii) the sharp spectral break at the peak 
frequency, inconsistent with the broad synchrotron bump, (iii) the evidence 
for two spectral sub-components, and (iv) possible existence of emission 
features called ``GRB lines''. We believe these facts strongly support both
the existence of small-scale magnetic fields and the proposed radiation 
mechanism from GRB shocks. As an example, we use the composite model to 
analyze GRB~910503 which has two spectral peaks. At last, we emphasize that 
accurate GRB spectra may allow precise determination of fireball properties
as early as several minutes after the explosion. 
\end{abstract}

\section{Introduction}

The conventional paradigm of the generation of radiation by relativistic 
electrons in magnetic fields is totally based on the theory of synchrotron 
radiation. We demonstrate that this theory is invalid if a magnetic field 
is tangled on very short spatial scales and we develop a quantitative theory 
of radiation in this case. Apparently, the required short-scale fields may 
naturally be present in astrophysical shocks. Here we focus on radiation 
from gamma-ray burst (GRB) shocks, for which a detailed theory of the 
formation of magnetic fields has recently been elaborated. 

The relativistic blast-wave model of cosmological GRBs 
\citep[see, e.g., a review by][]{Pi99} explains fairly well many 
observational features of this phenomenon, such as the rapid 
variability of $\gamma$-ray flux, the prompt optical flash, 
the light curves and spectra of delayed afterglows, etc..
This model interprets the prompt $\gamma$-ray flash as synchrotron 
radiation emitted by Fermi-accelerated electrons
in internal shocks propagating in the ejecta \citep{RM94} 
and then Lorentz-boosted to the $\gamma$-ray band. The afterglows
are explained in a similar way, as the emission from an external shock
\citep{MR93} propagating into the interstellar medium. To achieve the
observed very high luminosities, the magnetic field in the GRB shocks 
must be of nearly equipartition strength, $\epsilon_B=B^2/8\pi e_t\sim1$, 
where $e_t$ is the thermal energy density of the shocked material.

Until very recently, the equipartition assumption was completely unjustified.
\citet{ML99} have shown that the relativistic two-stream instability 
is capable of producing magnetic fields with $\epsilon_B\sim10^{-1}-10^{-5}$
in both internal and external shocks. Observations of afterglow spectra and 
light curves yield values of $\epsilon_B$ from $\sim10^{-1}-10^{-2}$ for
GRB~970508 \citep{WG98,FWK99,GPS99} to $\sim10^{-5}$ for 
GRB~990123 and GRB~971214 \citep{Galama99}. Recent detection of polarization 
of the optical afterglow of GRB~990510 \citep{Covino99,Wijers99}
indicates that the geometry of the magnetic field in the shock is consistent 
with the predictions of \citet{ML99} for collimated outflows 
\citep{GL99,Gruz99,Sari99}.

The magnetic field produced in GRB shocks randomly fluctuates on a very
small scale of roughly the relativistic skin depth, which is
$\lambda_B\sim10^2$~cm in internal shocks, for instance.
On the other hand, the emitting ultrarelativistic electrons
have much larger Larmor radii. Therefore, the electron trajectories are
not helical, as they would be in a homogeneous field. Thus, the theory of 
synchrotron radiation derived for homogeneous fields is not applicable
and the spectrum of the emergent radiation is different. Such a situation 
has never been considered in the astrophysical literature.

In this paper we investigate the effect of small-scale magnetic fields
on the properties of radiation. We primarily focus on internal shocks,
for concreteness. We show that there are two regimes, depending on the 
ratio of particle's deflection angle and the relativistic beaming angle.
Which regime is realized depends on the magnetic field properties, 
$B$ and $\lambda_B$, but is independent of the particle's energy.
When deflections are large compared to beaming, synchrotron radiation
is emitted. Otherwise, when particle's deflections are small, a new 
type of radiation --- jitter radiation --- is produced. A quantitative
analytical theory of this radiation is developed in this paper. For the 
power-law distributed electrons with a cutoff, $N(\gamma)\propto\gamma^{-p}$
for $\gamma\ge\gamma_{\rm min}$, where $\gamma$ is the particle's Lorentz 
factor and $p$ is the index, the emergent spectrum has the following 
properties. First, the spectral power peaks at the so-called jitter 
frequency, $\omega_{jm}=\omega_j(\gamma_{\rm min})$, which, 
unlike the synchrotron frequency, is independent of the magnetic 
field strength but, instead, depends on the particle density in the 
shock, $\omega_{jm}\propto\sqrt{n}$. Second, at low frequencies, 
$\omega\le\omega_{jm}$, the spectral power scales as 
$P(\omega)\propto\omega^1$, in contrast to the synchrotron spectrum,
for which $P(\omega)\propto\omega^{1/3}$. The high-frequency portion
is, however, determined by the electrons, $P(\omega)\propto\omega^{-(p-1)/2}$.
Third, the total (i.e., frequency integrated) powers emitted in the jitter and 
synchrotron regimes are identical.

Since large-scale fields (e.g., due to a magnetized progenitor, for instance) 
may also be present in the shocked material, we construct a composite, 
two-component jitter+synchrotron (JS) spectral model of the prompt $\gamma$-ray 
emission. We then compare the predictions of this model with presently 
available data collected (mostly) by the Burst And Transient Source Experiment 
(BATSE) on the {\it Compton Gamma-Ray Observatory} ({\it CGRO}). It turns out 
that the JS model is able to naturally explain some properties of the GRB 
spectra which are inconsistent with a simple synchrotron shock model. 
First, almost a half of all BATSE bursts violate the so-called
synchrotron ``line of death'' prediction, i.e., their low-frequency spectral
indices are greater than the maximal admissible value of $1/3$ \citep{Preece98}.
The spectra of these bursts are, however, well consistent with the steeper
$\omega^1$-law of jitter radiation. Second, the sharp change of the spectral 
index at a peak frequency seen in some bursts is also consistent with our model.
Third, the JS model theoretically supports
the fact that some GRBs have two spectral sub-components \citep{Pendl94}.
Fourth, there is an indication that spectra of some GRBs exhibit spectral
features --- ``GRB lines'' \citetext{see \citealp{Briggs99} for a review 
and some BATSE candidates}. It should be noted, however, that the results
from {\it Ginga} and BATSE are somewhat controversial. We demonstrate that
a line-like spectral feature may be associated with a weak jitter 
component in a synchrotron-dominated spectrum. If a future analysis shall 
confirm that ``GRB lines'' are real features and not instrumental 
(or other) artifacts, then they provide, together with the 
synchrotron component, precise information about properties of cosmological 
fireballs just a few hundred seconds after the explosion. We illustrate this 
on the example of GRB~910503 which has been observed by all four instruments 
on the {\em CGRO} and which exhibits a second spectral peak at roughly twice
the synchrotron peak frequency. A simple fit of a spectral shape readily 
yields the value of the magnetic field, $\epsilon_B\sim4\times10^{-4}$ in 
the shock, which is in agreement with results  of a completely different 
analysis by \citet{Dermer99}. It should be noted that a reliable 
identification/detection of the jitter spectral features will provide 
a direct evidence that the magnetic fields in GRBs are generated by 
the two-stream instability, since this is the only presently known mechanism 
capable of producing small-scale, large-amplitude fields in shocks.

At last, we emphasize that the advantage of our JS model is that it was 
{\em not} specially designed in order to explain peculiarities of the GRB
spectra, but solely to study the physical effect of small-scale magnetic fields.
The phenomenon considered in this paper is quite general and, clearly,
relevant not only to the emission from internal GRB shocks. A similar mechanism
of emission is expected to operate in external GRB shocks  and, possibly, in 
more conventional supernova shocks and blazar jets. 

The rest of this paper is organized as follows. A qualitative consideration
of the radiation mechanisms in a nonuniform magnetic field is presented in
\S \ref{S:GC}. In \S \ref{S:M} the structure and properties of magnetic 
fields in GRB internal shocks are discussed. In \S \ref{S:QT} we present
a quantitative  analytical theory of jitter radiation. A two-component,
jitter+synchrotron spectral model of the prompt $\gamma$-ray emission
is presented in \S \ref{S:JS}. We compare the predictions of the model with
recent observational results in \S \ref{S:D}. 
Finally, \S \ref{S:C} is the conclusion.

\section{Radiation from Small-Scale Fields: General Consideration
\label{S:GC} }

Let's consider a plasma at rest threaded by a magnetic field. Let's now
consider an ultrarelativistic electron with the Lorentz factor
$\gamma\gg1$ moving in a magnetic field. Because of beaming the emerging 
radiation is concentrated in a narrow cone with the opening angle 
$\Delta\theta\sim1/\gamma\ll1$ in the direction of the particle's velocity.
In a uniform magnetic field the electron moves along a helical trajectory,
so that the radiation seen by an observer consists of short pulses repeated 
every cyclotron period. The synchrotron spectrum, therefore, consists of a
large number of cyclotron harmonics, the envelope of which is determined by the 
inverse duration of the pulses \citep{RL}. The spectrum is peaked near the
critical synchrotron frequency $\omega_c=\frac{3}{2}\gamma^2eB_\bot/m_ec$,
where $B_\bot=B\cos\chi$ and $\chi$ is the particle's pitch angle.

If the magnetic field is randomly tangled and the correlation length is less 
then a Larmor radius of an emitting electron, then the electron experiences
random deflections as it moves through the field. Its trajectory is,
in general, stochastic. This is similar to a collisional motion of 
an electron in a medium. Bremsstrahlung quanta are emitted in every collision.
Unlike the bremsstrahlung case, here ``collisions'' are due to small-scale 
inhomogeneities of the magnetic field rather than due to electrostatic fields 
of other charged particles. Since the Lorentz force depends on particle's
velocity, the emergent spectrum will be somewhat different from pure 
bremsstrahlung.
There is also an alternative physical interpretation of the process.
For an ultrarelativistic electron, the method of virtual quanta applies
\citep{RL}. In the rest frame of the electron, the magnetic field inhomogeneity
with wavenumber $k$ is transformed into a transverse pulse of electromagnetic 
radiation with frequency $kc$. This radiation is then Compton scattered by 
the electron to produce observed radiation with frequency $\sim\gamma^2kc$ 
in the lab frame.  

Keeping this general physical picture in mind, we now analyze the problem
in more details.
Let's  consider a nonuniform random magnetic field with a typical 
correlation scale $\lambda_B$, the Larmor radius of the electron,
$\rho_e=\gamma m_ec^2/eB_\bot$ is less or comparable  comparable to 
$\lambda_B$. The emerging spectrum depends on the relation between the
particle's deflection angle, $\alpha$, and the beaming angle, $\Delta\theta$, 
\citep{LL}. For ultrarelativistic 
particles and small deflection angles, the latter is estimated as follows. 
The particle's momentum is $p\sim\gamma m_ec$. The change in the 
perpendicular momentum due to the Lorentz force acting on the particle during
the transit time $t\sim\lambda_B/c$ is $p_\bot\sim F_Lt\sim eB_\bot\lambda_B/c$.
The angle $\alpha$ is then $\alpha\sim p_\bot/p\sim 
eB_\bot\lambda_B/\gamma m_ec^2$. Thus, the ratio of the two angles is 
\beq
\frac{\alpha}{\Delta\theta}\sim\frac{eB_\bot\lambda_B}{m_ec^2}
\sim\gamma\frac{\lambda_B}{\rho_e} .
\eeq
It is interesting to note that this ratio is independent of particle's 
energy (i.e., of $\gamma$) and is determined by the properties of the magnetic 
field only, i.e., by $B$ and $\lambda_B$. It is more convenient, however, to 
use the wave-vector, $k_B$, as a measure of the magnetic field scale, instead 
of $\lambda_B\sim k_B^{-1}$. We now define the deflection-to-beaming ratio
as follows,
\beq
\delta\equiv\frac{\gamma}{k_B\rho_e}\sim\frac{\alpha}{\Delta\theta}.
\label{delta}
\eeq
There are two limiting cases.

\bfig
\plotone{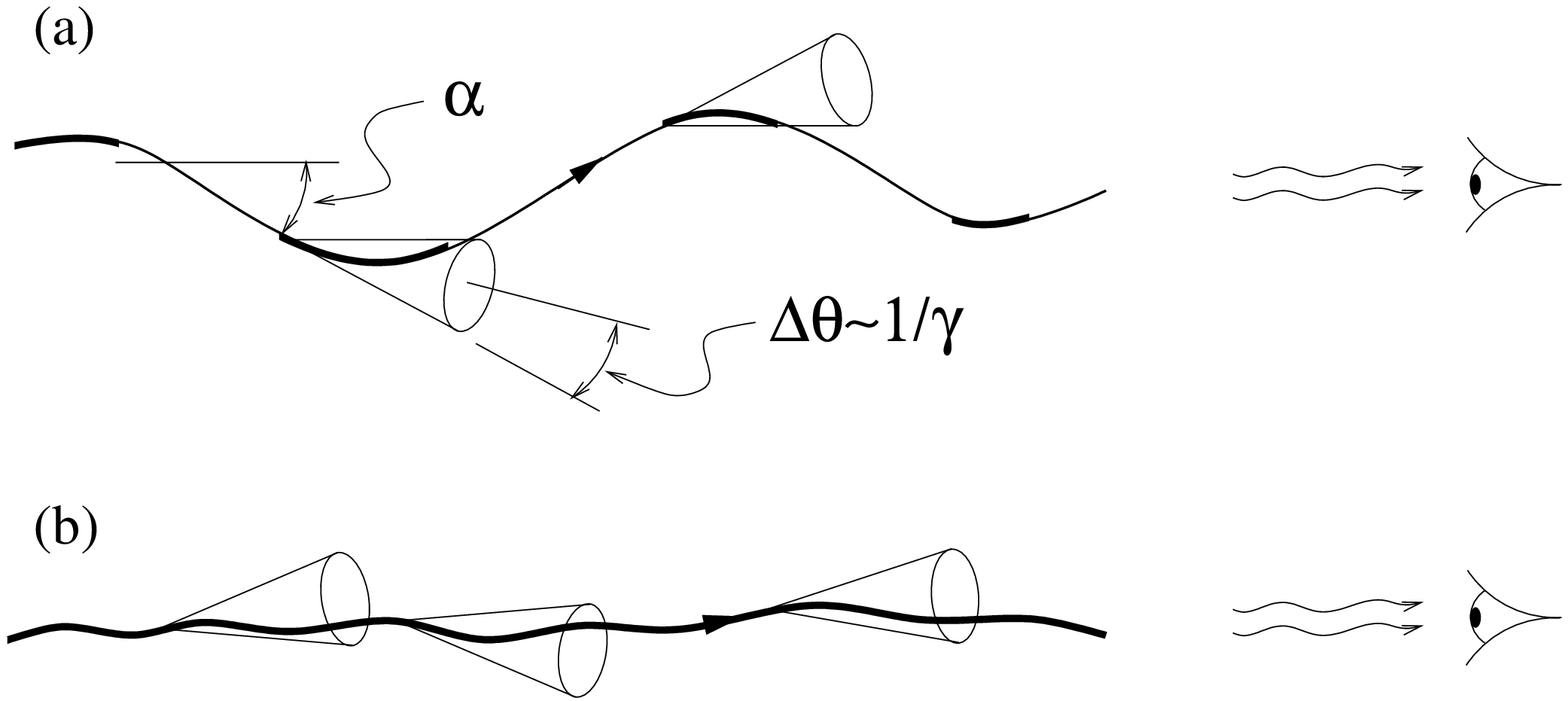}
\figcaption[paths.ps]{Emission from various points along  particle's 
trajectory; 
(a) --- $\alpha\gg\Delta\theta$, emission from selected parts (bold portions) 
of the trajectory is seen by an observer, 
(b) --- $\alpha\ll\Delta\theta$, emission from the entire trajectory 
is observed.
\label{fig1} }
\efig

First, $\delta\sim\alpha/\Delta\theta\gg1$, i.e., the deflection angle 
is much larger than the beaming angle, see Figure \ref{fig1}a. Then, an 
observer sees radiation coming from short segments (``patches'') 
of the electron's trajectory which are nearly parallel to the line of 
sight (very much like for pure synchrotron radiation). The magnetic field in 
every patch is almost uniform but it varies from patch to patch. The radiation
is pulsed with a typical duration $\tau_p\sim 1/\omega_c$. The characteristic 
frequency of the observed radiation is thus $\omega\sim\omega_c$.  
Note that in the time- or ensemble-averaged spectrum, 
the instantaneous field, $B_\bot$, which enters $\omega_c$, 
should be appropriately averaged, $\bar B\simeq\sqrt{\langle B^2_\bot\rangle}$.
In this case the emergent radiation is completely identical to 
synchrotron radiation from large-scale weakly inhomogeneous magnetic fields.

Second, $\delta\sim\alpha/\Delta\theta\ll1$, i.e., the deflection 
angle is smaller than the beaming angle, so that the entire electron's 
trajectory is seen by an observer, as shown in Figure \ref{fig1}b. 
The particle moves along the line of sight almost straight and 
experiences high-frequency jittering in the perpendicular direction 
due to the random Lorentz force. We therefore refer the emerging radiation
to as ``jitter'' radiation. Its spectrum is determined by random 
accelerations of the particle. Let's imagine an electron moving 
ultrarelativistically along the line of sight with a constant velocity, 
the transverse accelerations of the electron are small.
In the laboratory frame, the electron passes through the magnetic field 
inhomogeneities having a typical scale $\lambda_B\sim k_B^{-1}$ with the 
velocity $c$. In the particle's frame 
(i.e., where its parallel velocity vanishes), the field correlation 
scale is $\lambda'_B\sim\lambda_B/\gamma\sim(k_B\gamma)^{-1}$ 
due to the Lorentz transformation. The electron's perpendicular acceleration 
changes significantly during $\tau\sim\lambda'_B/c\sim(\gamma k_Bc)^{-1}$,
so that the characteristic frequency of the emitted radiation is 
$\omega_{j0}\sim\tau^{-1}\sim \gamma k_Bc$. In the laboratory frame, this
frequency is boosted to $\omega_j=\gamma\omega_{j0}$. Thus, the spectrum of the 
emergent radiation is peaked at the frequency $\omega_j\sim\gamma^2k_Bc$.
This frequency is higher than the critical synchrotron frequency in the 
uniform magnetic field of the same strength, $\omega_{c*}$, namely
\beq
\frac{\omega_{c*}}{\omega_j}\simeq\frac{3}{2}\,\delta\ll1 ,
\label{wcwj}
\eeq
as follows from equation (\ref{delta}). 

We should warn here that despite apparent similarity of the jitter
and free electron laser emission mechanisms, they are quite different.
The wiggler field in free electron lasers is appropriately adjusted for
the electron motion to be in phase with the produced radiation field
to emit coherent radiation. Jitter radiation is, in general, incoherent.

\section{The structure of the magnetic field in GRB shocks
\label{S:M} }

To proceed further, a model for a magnetic field in GRB shocks is required. 
We use the only presently available quantitative theory of the magnetic
field generation in shocks proposed by \citet{ML99}. To be specific, we focus 
on internal shocks which produce $\gamma$-ray emission. External shocks which 
are responsible for the delayed afterglows may be treated similarly and will 
be considered in a future publication. 

Shock fronts are shown to be natural sites of the magnetic 
field generation. Right before a shock, the inflowing (in the shock frame) bulk
plasma particles meet the outflowing particles which were reflected (scattered) 
from the shock. Such a two-stream motion is kinetically unstable. 
The emergent magnetic field is random with zero mean and lies in the plane 
of the shock front, --- perpendicular to the shock velocity. In principle,
all plasma species participate in the instability. We assume the protons and 
electrons to be the only species and discuss their contributions separately.

It is important to emphasize that the generated magnetic field fills 
the entire volume of a shock shell and is not located within a thin layer 
of order several skin depths near the front. There is a gas flow though a
shock. Because of flux freezing the generated magnetic field is transported 
with the shocked material downstream. This material is replenished with
a fresh one where a new magnetic filed is thus continuously produced.
Since the magnetic field is long-lived and does not decay in a dynamical time,
as indicated by numerical simulations \citetext{see references in 
\citealt{ML99}}, this field will be present in the entire ejecta.

\subsection{Fields produced by the electrons and protons \label{S:MEP}}

In this subsection we briefly remind main results of the theory of
\citet{ML99} for future reference.
Since electrons are light, the instability induced by them is rapid: 
the typical $e$-folding length (i.e., the $e$-folding time times the 
shock speed) is much smaller than the characteristic shock thickness
determined by the Larmor radius of heavier protons. 
Therefore, the magnetic field energy grows rapidly and reaches 
the approximate equipartition with the electron kinetic energy,
\beq
\frac{\bar B^2_e}{8\pi}=\eta_e\gamma_{\rm int} m_ec^2n
=\frac{m_e}{m_p}\,e_t\eta_e, 
\label{Be}
\eeq
i.e., $\epsilon_{Be}=\bar B_e^2/8\pi e_t=(m_e/m_p)\eta_e$,
where $\eta_e$ is the efficiency factor for the electrons which incorporates 
uncertainties due to the nonlinear phase of the instability, one infers 
from numerical simulations that generically $\eta_e\simeq0.1-0.01$, 
$\gamma_{\rm int}$ is the relative Lorentz factor of two colliding shells which 
produce an internal shock, $\gamma_{\rm int}\sim2-4$, and $n$ is the  
number density of particles in the expanding shell, before the shock. 
The spatial correlation scale, $k_{Be}$, of the field behind the shock is 
\beq
k_{Be}=\frac{4\gamma_{\rm int}\omega_{pe}}{2^{1/4}\bar\gamma_e^{1/2}c} ,
\label{kBe}
\eeq
where $\omega_{pe}^2=4\pi e^2 n/m_e$ is the electron plasma frequency squared, 
$\bar\gamma_e$ is the initial effective thermal Lorentz factor of the 
streaming electrons, and an extra factor of $4\gamma_{\rm int}$ is due to the 
relativistic shock compression. 

The protons may generate magnetic fields too.
Since they are much heavier than electrons, the spatial coherence length
and the $e$-folding length are comparable to 
the thickness of the collisionless shock. Therefore, the field does not have 
enough time to grow during the flow transit through the shock. It may however 
grow behind the shock if the two-stream motion of the protons persists there.
If this is the case, then there are two possibilities. First, if there is no 
energy transfer from the protons to the electrons or if it is slow 
compared to the rate of the field growth, then the magnetic field energy may 
be as large as
\beq
\frac{\bar B_p^2}{8\pi}=\eta_p\gamma_{\rm int}m_pc^2n=e_t\eta_p\sim0.1e_t , 
\eeq
provided $\eta_p\sim0.1$. Second, if the energy transfer is fast, which may
be the case in fields which are in equipartition with the electrons or 
stronger, then the protons may efficiently damp their energy 
into the emitting electrons, so that the resultant field will be 
$\bar B_p^2/8\pi\sim\bar B_e^2/8\pi$. Alternatively, if no magnetic 
field is generated downstream the shock, the strength of the field 
produced by the protons may be orders of magnitude lower. Which case 
realizes may be learned from numerical particle simulations or from 
observations. We keep $\epsilon_{Bp}=\bar B^2_p/8\pi e_t$ as a parameter.
The characteristic correlation scale of the generated magnetic field is
\beq
k_{Bp}=\frac{\omega_{pp}}{2^{1/4}\bar\gamma_p^{1/2}c} ,
\eeq
where $\omega_{pp}^2=4\pi e^2 n/m_p$, $\bar\gamma_p\sim2$ is the initial 
effective thermal Lorentz factor of the streaming protons. The compression 
factor, $4\gamma_{\rm int}$, is absent because the field is likely produced 
after the compression occurs.

\subsection{The model \label{S:MMODEL}}

From equations (\ref{delta}), (\ref{Be}), and (\ref{kBe}), we
estimate the $\delta$-parameter for the electrons,
\beq
\delta_e=\frac{1}{2^{7/4}}\left(\frac{\bar\gamma_e}{\gamma_{\rm int}}
\right)^{1/2}\sqrt{\eta_e}\equiv\phi\sqrt{\eta_e}\lesssim1.
\label{delta-e}
\eeq
The exact value of $\phi=2^{-7/4}(\bar\gamma_e/\gamma_{\rm int})^{1/2}$ 
is somewhat uncertain: $\bar\gamma_e$ may evolve during the instability 
from its initial value $\bar\gamma_e\sim 2-3$ to 
$\bar\gamma_e\sim\gamma_{\rm int}\sim3-4$ due to nonlinear effects.%
\footnote{Note, the inflowing electrons are cold; they are not Fermi
accelerated yet. Note also that $\bar\gamma_e$ cannot be greater than 
$\gamma_{\rm int}$ for the instability to operate. Otherwise, no magnetic 
field is produced.}
The numerical prefactor may also be affected. Therefore we assume that 
possible values of $\delta_e$ are in the range 
$1\lesssim\delta_e\lesssim10^{-2}$ (given the uncertainty in $\eta_e$ 
from 0.1 to 0.01) and generically $\delta_e\sim0.1$. 
The $\delta$-parameter for the protons is 
\beq
\delta_p=2^{-1/4}\left(\bar\gamma_p\gamma_{\rm int}\right)^{1/2}
\frac{m_p}{m_e}\,\sqrt{\eta_p}\gg1,
\eeq
unless $\eta_p$ is too small: $\eta_p\lesssim10^{-7}$ 
(i.e., $\epsilon_{Bp}\lesssim10^{-3}\epsilon_{Be}$).

As will be shown below in \S \ref{S:QT}, a spatial spectrum of the magnetic 
field $\bar B_e$ with $\delta_e<1$ is required to calculate the spectrum of 
jitter radiation.
This distribution of $\bar B_e$ over scales is difficult to find from the 
first principles because it is determined by fully nonlinear dynamics of
the instability process. Some constraints may however be drawn. 
First, the two-stream instability 
produces magnetic fields in a finite range of scales,
$0\le k\le k_{{\rm crit},e}$, where $k_{{\rm crit},e}\sim k_{Be}$ 
to within a numerical factor of order unity.
Second, the field grows until it becomes 
strong enough to deflect the particles in the transverse direction by
$\sim 1$ radian on a scale of the field coherence length, i.e.,
$\bar B_e^{-1}\propto\rho_e\sim k_{Be}^{-1}$. If we now assume 
that each Fourier harmonic, $B_k$, is amplified independently of others,
we obtain: $B_k\propto k$ for $k\le k_{{\rm crit},e}$. This is the lower 
limit: the spectrum of the magnetic field can only be steeper than linear.

In reality, all Fourier harmonics are coupled. Thus, when at least one 
harmonic reaches the sub-equipartition strength, the streaming electrons are 
isotropized by random Lorentz forces. This prevents the growth of other 
harmonics. Therefore, the spectrum of the field will be steeper than 
linear and will have a maximum near $k_{Be}$. 
The simplest possible model is a power-law,
\beq
B_k=\left\{\begin{array}{ll}
C_Bk^\mu, & \textrm{ if $0\le k\le k_{Be}$}, \\
0, 	  & \textrm{ otherwise} ,
\end{array}\right.
\label{Bk}
\eeq
where $C_B$ is a normalization constant and $\mu\ge1$ is a power-law spectral 
index of the magnetic field being a free parameter here. The constant $C_B$
may be determined using Parseval's theorem,%
\footnote{We use the following definition of the Fourier transform:
$f(\omega)=\int f(t)e^{-i\omega t}dt$. }
$\lim_{L\to\infty}\int_{-L/2}^{L/2}B^2_e\,dx
=\frac{1}{\pi}\int_0^\infty B_k^2\,dk$, where $L$ is the system size.
Taking into account that $\int B^2_e\,dx=\bar B_e^2L=\bar B^2_e cT$, 
where $T$ is the total duration of the pulse, we write
\beq
C_B^2=\pi\left(2\mu+1\right)cT{\bar B^2_e}{k_{Be}^{-(2\mu+1)}} .
\label{C}
\eeq
The magnitude of the field, $\bar B_e$, is given by equation (\ref{Be}).

Strictly speaking, this model may be used for not too small $k$'s: 
$k>k_{\rm min}$, where $k_{\rm min}$ is set by the condition: 
$\gamma/(\rho_e k_{\rm min})=1$, see equation (\ref{delta}). 
The field harmonics with $k<k_{\rm min}$ are large-scale ones and
they contribute to synchrotron radiation. Using equation (\ref{delta}), 
it is easy to obtain that $k_{\rm min}=\delta_e\,k_{Be}$.
It is also useful to introduce the small-scale field component
as follows, 
\beq
\bar B^2_{SS}=\int_{k_{\rm min}}^{k_{Be}}B_k^2\,dk,
\qquad \frac{\bar B^2_{SS}}{\bar B^2_e}=1-\delta^{2\mu+1}.
\label{Bss}
\eeq
Hereafter, we omit the subscripts by $\delta$; 
this should not cause any confusion.

\bfig
\plotone{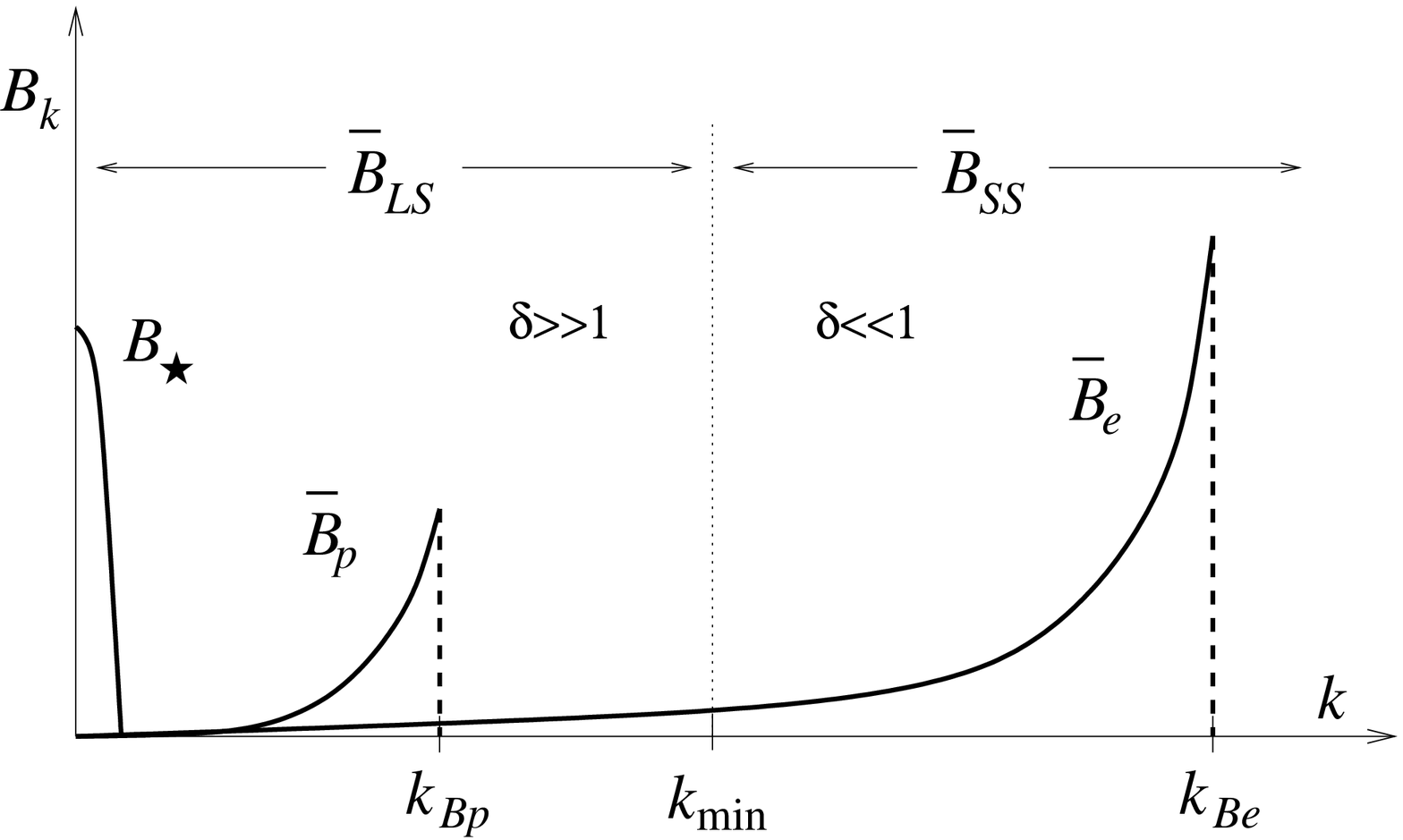}
\figcaption[magnfield.eps]{ The model for the magnetic field in GRB shocks.
\label{fig:magnfield} }
\efig

To summarize, we assume the following structure of the magnetic
field sketched in Figure \ref{fig:magnfield}. First, there is a magnetic field 
produced by the electrons, $\bar B_e$, the magnitude of which is given by 
equation (\ref{Be}). A large fraction of this field contributes to the 
small-scale component, in accordance with equation (\ref{Bss}). Its 
distribution over spatial scales is described by a power-law with the 
index $\mu$, see equation (\ref{Bk}). The rest, $\bar B^2_e-\bar B_{SS}^2$, 
contributes to the large-scale component. Second, there is a magnetic field 
produced by the protons, $\bar B_p$. This field is a large-scale field.
Both $\bar B_e$ and $\bar B_p$ are random with zero mean.
Third, there could be an ordered magnetic field left from a magnetized 
progenitor, $B_{\star}$. We define the total large-scale magnetic field
as follows, 
\beq
\bar B_{LS}^2=\bar B_0^2+\left(\bar B_e^2-\bar B_{SS}^2\right), \qquad
\bar B_0^2=B_{\star}^2+\bar B_p^2 ,
\label{Bls}
\eeq
where $\bar B_0$ is the fraction of the large-scale field which is not 
produced by the electrons.

\section{Radiation from Small-Scale Fields: Quantitative Theory
\label{S:QT} }

In \S \ref{S:GC} we qualitatively demonstrated that there are two regimes of 
radiation. An ultrarelativistic electron propagating through 
small-scale fields generated only by the electrons ($\delta\ll1$) 
emits jitter radiation. The electron moving through larger-scale
fields generated only by the protons ($\delta\gg1$) emits synchrotron radiation.
The radiation spectrum from a magnetic field with a broadband distribution over 
scales is neither of the above and must be calculated by appropriate scale 
averaging of a particle trajectory. However, for a bimodal field distribution
of \S \ref{S:MMODEL} such that the magnetic energy in $B_k$ harmonics for
which $\delta\sim 1$ (i.e., where the transition from a jitter to synchrotron 
regime occurs) is small, the separation of scales is possible. The resultant 
radiation will approximately be a composition of the jitter and synchrotron
spectra.\footnote{Physically, a particle moves along a helical trajectory
	about field lines of a large-scale field. This trajectory is slightly 
	perturbed by high frequency jittering in the (instantaneous) 
	transverse direction due to a small-scale field component. 
	The intensities of the spectral sub-components are determined
	by the magnetic field energy densities at large and small scales.}
For clarity and illustration purposes, we derive the radiation properties and 
spectra in both limits separately. A rigorous treatment of a general case will 
be given elsewhere. All calculations in this section are done in the reference 
frame of the shocked material, unless stated otherwise.

\subsection{Jitter radiation, \(\delta\ll1\) }

For sufficiently small $\delta$'s such that $\alpha\ll\Delta\theta$, 
the velocity ${\bf v}$ of a particle is almost constant whereas its 
acceleration ${\bf w\equiv\dot v}$ varies with time. Calculating the Fourier 
component of the electric field using the Li\'enard-Wiechart (retarded) 
potentials, one arrives at the following expression for the total energy 
emitted per unit solid angle $d\Omega$ per unit frequency $d\omega$ 
\citetext{\citealp{LL}, \S 77; see also \citealp{RL}, \S 3.2}:
\beq
dW=\frac{e^2}{2\pi c^3}\left(\frac{\omega}{\omega'}\right)^4
\left|{\bf n}\times\left[\left({\bf n}-\frac{\bf v}{c}\right)\times
{\bf w}_{\omega'}\right]\right|^2 d\Omega\,\frac{d\omega}{2\pi},
\label{dW-1}
\eeq
where ${\bf w}_{\omega'}=\int{\bf w}e^{i\omega't}\,dt$ is the Fourier 
component of the particle's acceleration, 
$\omega'=\omega\left(1-{\bf n\cdot v}/c\right)$, and ${\bf n}$ is
the unit vector pointing towards the observer.

First, we can simplify the vector expression in (\ref{dW-1}).
Indeed, in the ultrarelativistic case, the longitudinal component of the
acceleration is small compared to the transverse component, 
$w_\|/w_\bot\sim1/\gamma^2\ll1$. Therefore ${\bf v}$ and ${\bf w}$ are
approximately perpendicular to each other. Second, the dominant 
contribution to the integral over $d\Omega$ comes from small angles 
$\theta\sim1/\gamma$ with respect to the particle's velocity. Therefore, 
we approximately write $\omega'\simeq\omega\left(1-v/c+\theta^2/2\right)
\simeq\frac{1}{2}\omega\left(1-v^2/c^2+\theta^2\right)
=\frac{1}{2}\omega\left(\theta^2+\gamma^{-2}\right)$. We now can replace
integration over the solid angle $d\Omega\simeq\theta\,d\theta\,d\phi$ with
integration over $d\phi\,d\omega'/\omega$ and integrate equation (\ref{dW-1})
over the azimuthal angle, $\phi$, from $0$ to $2\pi$.
The spectral energy finally becomes 
\beq
\frac{dW}{d\omega}=\frac{e^2\omega}{2\pi c^3}\int_{\omega/2\gamma^2}^\infty
\frac{\left|{\bf w}_{\omega'}\right|^2}{\omega'^2}
\left(1-\frac{\omega}{\omega'\gamma^2}+\frac{\omega^2}{2\omega'^2\gamma^4}
\right)\,d\omega' .
\label{dW/dw}
\eeq
The leading term in the above equation is due to high-frequency ``jittering'' 
of the electron as it moves through the random magnetic field. The second 
and third terms in the brackets are corrections due to the angular structure 
of the radiation field convolved with the relativistic beaming.

The acceleration ${\bf w}$ is found from the equation of motion,
$\dot{\bf p}=(e/c){\bf v\times B}$. In general, ${\bf B}$ may vary both 
in amplitude and in direction. In relativistic GRB shocks, radiation is 
beamed; only a conical section bound by the opening angle 
$\theta_b\sim1/\gamma_{\rm sh}$ is seen by an observer, where $\gamma_{\rm sh}$ 
is the bulk Lorentz factor of the expanding shell. At a particular 
observing time, most of the radiation is coming from the brightened limb
\citep{Sari98,PM98}. Because of relativistic aberration \citep{RL}, 
the shock front is seen almost edge-on at the limb. On the other hand, 
the magnetic fields in the shock shell are random but always lie in the 
plane of the shock front \citep{ML99}. Choosing a coordinate system
with $\hat x$-direction pointing towards the observer (in the shock frame) and 
$\hat x$-$\hat y$-plane being the shock front plane at the limb, we have
${\bf v}=c\hat x$, ${\bf B}_\bot=B\hat y$, and 
$\dot{\bf p}\simeq\gamma m_ew\hat z$. Then, the Fourier component of $w$ is
\beq
w_{\omega'}=\frac{eB_{\omega'}}{\gamma m_e}=\frac{eB_{k'}}{\gamma m_ec}.
\label{w-omega}
\eeq
For the magnetic field distribution, we use the model (\ref{Bk}).

At last, the spectral power $P(\omega)\equiv dW/dt\,d\omega$ 
(i.e., the energy emitted per unit time per unit frequency interval, 
measured in ergs~s$^{-1}$~Hz$^{-1}$) is obtained 
dividing the energy spectrum $dW/d\omega$ by the pulse duration, 
$T$ \citep{RL}. Combining equations (\ref{w-omega}), (\ref{Bk}), (\ref{C}) 
together and substituting into (\ref{dW/dw}), we obtain
\beq
P(\omega)=r_e^2c\gamma^2
\frac{\bar B^2_{SS}}{2\omega_j}\;J\!\left(\frac{\omega}{\omega_j}\right) ,
\label{Pr(w)}
\eeq
where $\omega/\omega_j\le2$, and $r_e=e^2/m_ec^2$ is the classical 
electron radius. As is expected, the characteristic frequency of the
 emergent radiation is 
\beq
\omega_j=\gamma^2k_{{Be}}c
=2^{7/4}\gamma^2\gamma_{\rm int}\bar\gamma_e^{-1/2}\omega_{pe},
\label{omegaj}
\eeq
cf., equations (\ref{kBe}) and (\ref{delta-e}). The function $J$ is defined as
\beq
J(\xi)=(2\mu+1)\xi^{2\mu}\left[I\!\left(\textrm{min}\!
\left[2;\,\frac{\xi}{\delta}\right]\right)-I(\xi)\right], 
\eeq
where $I$ is the integral,%
\footnote{ We keep this integral in a general form because it contains 
logarithmic terms for $\mu=0.5,\ 1,\ 1.5$.}
$I(\xi)=\int\xi^{-2\mu}\left(1-\xi+\frac{1}{2}\xi^2\right)\,d\xi$,
and $\textrm{min}[a;b]$ denotes the smallest of $a$ and $b$. 
From equations (\ref{delta}), (\ref{Be}), (\ref{omegaj}), we have
$\omega_j=\gamma^2\bar B_e/\delta m_e c$.
Equation (\ref{Pr(w)}) may now be cast into the form,
\beq
P(\omega)=\frac{e^2}{2c}\,\delta^2\,
\frac{\omega_j}{\gamma^2}\,\frac{\bar B^2_{SS}}{\bar B^2_e}\,
J\!\left(\frac{\omega}{\omega_j}\right) .
\label{Pr(w)1}
\eeq
This spectrum is shown in Figure \ref{fig:PrSingle} for several values 
of $\mu$. In general, the steeper the field distribution, $B_k\propto k^\mu$, 
the more peaked is the radiation spectrum and the closer the peak frequency to 
$2\omega_j$. Note, the discontinuity of the slope is artificial. This is 
due to the $k_{\rm min}$-cutoff because the model we use does not continuously
interpolate between the jitter and synchrotron radiation limits 
(i.e., between $\delta\ll1$ and $\delta\gg1$ cases).
This discontinuity is less prominent for large $\mu$'s and small $\delta$'s.

\bfig
\plotone{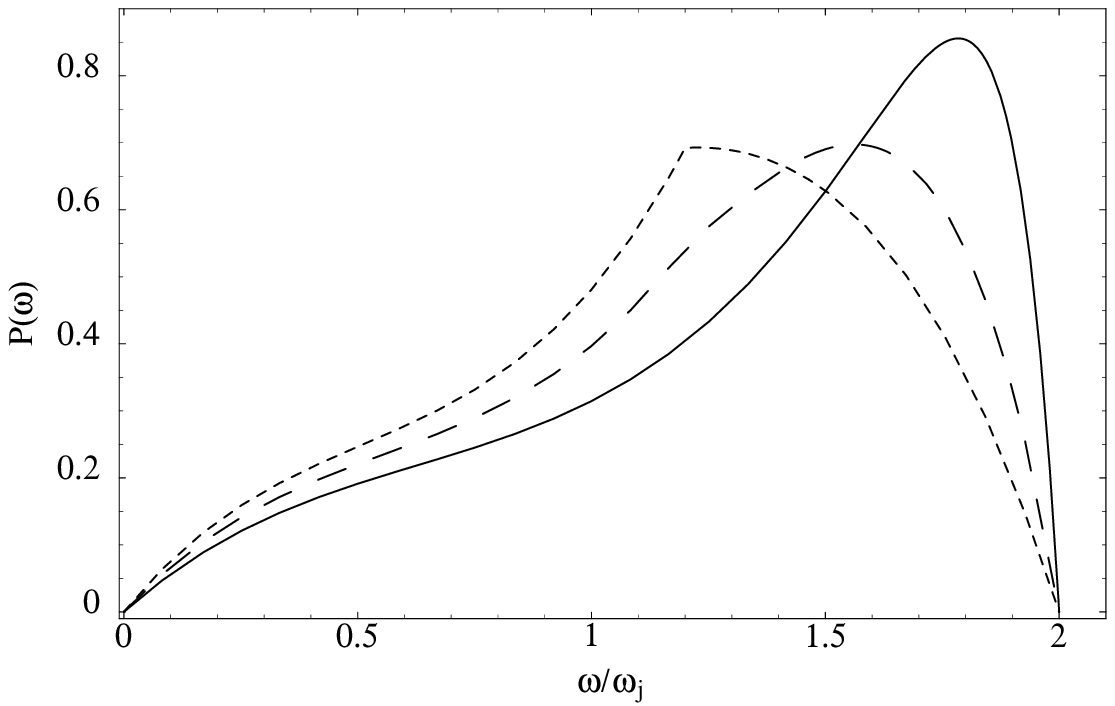}
\figcaption[PrSingle.eps]{ 
The arbitrarily normalized power spectra, $P(\omega)$, 
of jitter radiation (linear-linear plot) emitted by a single electron
for several values of the magnetic field spectral index, $\mu=1$ (dotted),
$\mu=3$ (dashed), and $\mu=10$ (solid).
Here $\delta=0.6$ and we keep the value of the small-scale magnetic field 
fixed, $\bar B^2_{SS}/8\pi=\bar B^2_e/8\pi$.
\label{fig:PrSingle} }
\efig

The total emitted power may be obtained by integrating (\ref{Pr(w)}) over 
frequencies. A simpler way is to express it in terms of electron's acceleration,
i.e., $dW/dt=(2e^2\gamma^4/3c^3)(w_\bot^2+\gamma^2w_\|^2)$ \citep{RL} and 
substitute $w_\|=0$ and $w_\bot$ given by (\ref{w-omega}). The result is
\beq
dW/dt=(2/3)r_e^2c\gamma^2\bar B^2_{SS}.
\eeq
Note, this expression is identical to that for a synchrotron radiation
in which a uniform field, $B_\bot^2$, is replaced with its average,
$\bar B^2_{SS}=\langle B^2_\bot\rangle$.

We are now able to calculate the jitter radiation spectrum from an
ensemble of electrons.
We assume that electrons are accelerated in the shock%
\footnote{Small-scale magnetic fields present in the shock provide effective
collisions in the otherwise collisionless plasma and make the Fermi acceleration
to operate \citep{ML99}.}
to a power-law distribution $N(\gamma)=C_N\gamma^{-p}$ (where $C_N$ is a
normalization constant) with a minimum Lorentz factor, 
i.e.,  $\gamma_{\rm min}\le\gamma<\infty$. The index $p$ must be 
$p\ge2$ so that energy does not diverge at large $\gamma$'s. 
We assume the standard value,%
\footnote{Resent studies indicate that the high-frequency spectral 
power index in the prompt GRBs is $\sim-1$ which translates to $p\sim3$. 
We keep $p=2.5$ for illustrative purposes.}
 $p=2.5$ \citep{SNP96}, unless stated otherwise.
The spectrum is found from 
\beq
P_{\rm ens}(\omega)=\int_{\gamma_{\rm min}}^\infty 
N(\gamma)P(\omega)\, d\gamma ,
\eeq
noting that $\omega_j\equiv\omega_j(\gamma)\propto\gamma^2$, i.e., in exact 
analogy with synchrotron radiation, $\omega_c\propto\gamma^2$. We obtain
\beq
P_{\rm ens}(\omega)=\frac{C_N}{2\gamma_{\rm min}^{p-1}}
\left(\frac{\omega}{\omega_{jm}}\right)^{-\frac{p-1}{2}}
\int_0^{{\omega}/{\omega_{jm}}}\xi^{(p-1)/2}\,P(\xi)\,d\xi ,
\label{Pens(w)}
\eeq
where we introduced the characteristic frequency of radiation
$\omega_{jm}\equiv\omega_j(\gamma_{\rm min})$, cf. equation (\ref{omegaj}),
\beq
\omega_{jm}
=2^{7/4}\gamma_{\rm min}^2\gamma_{\rm int}\bar\gamma_e^{-1/2}\omega_{pe},
\label{omegajm}
\eeq
At low and high frequencies, the spectra may be obtained analytically
as follows,
\beq
P_{\rm ens}(\omega)\propto
\left\{\begin{array}{ll}
\displaystyle{\left(\frac{\omega}{\omega_{jm}}\right)^1}, 
	& \omega\ll\omega_{jm}, \\
\displaystyle{\left(\frac{\omega}{\omega_{jm}}\right)^{-(p-1)/2}}, 
	& \omega\gtrsim\omega_{jm} ,
\end{array}\right.
\label{asymptotics}
\eeq
At high frequencies, the spectrum is analogous to the synchrotron case.
The electron power-law index determines the slope. At low frequencies, the
spectrum is linear in frequency, in contrast to the synchrotron spectrum
which is softer and scales as $P_{\rm syn}(\omega)\propto\omega^{1/3}$. 
The spectrum of radiation from single-speed electrons is peaked at 
$\omega\simeq2\omega_j$
for large $\mu$'s, i.e., for a peaked magnetic field distribution,
as discussed above (see also Figure \ref{fig:PrSingle}). Therefore, the
spectral break due to the $\gamma_{\rm min}$-cutoff occurs at
\bea
\nu_{j,{\rm break}}&\simeq&2\omega_{jm}/2\pi
=(2^{7/4}/\pi)\gamma^2_{\rm min}\gamma_{\rm int}\bar\gamma_e^{-1/2}\omega_{pe}
\nonumber\\
&\simeq&6.0\times10^9\,\gamma^2_{\rm min}\gamma_{\rm int}\bar\gamma_e^{-1/2}
n_{10}^{1/2}\textrm{ Hz} ,
\label{w-br}
\eea
where $n_{10}\equiv n/10^{10}\textrm{ cm}^{-3}$. The above frequency is
calculated in the frame of the relativistic expanding shell. This frequency
is boosted in observer's frame by a factor of $\gamma_{\rm sh}$.

The model of jitter radiation contains two extra parameters, compared to 
the model of synchrotron radiation. These parameters are the 
deflection-to-beaming ratio, $\delta$, and the magnetic field index, $\mu$, 
which determines the peakedness of the magnetic field over spatial scales. 
We now investigate how the spectrum of the emergent radiation depends on 
these parameters.

\bfig
\plotone{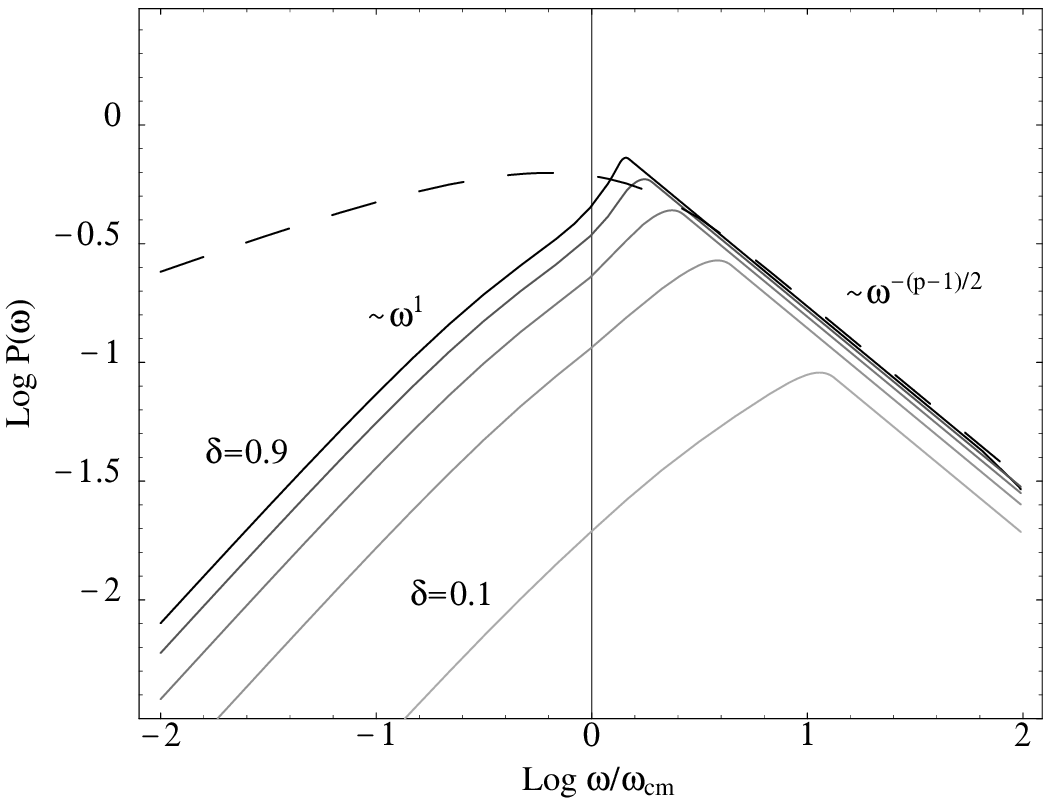}
\figcaption[PrEnsDelta.eps]{ 
Spectral power $P(\omega)$ of radiation (log-log plot, arbitrary units) 
emitted by the power-law distributed electrons vs. $\omega/\omega_{cm}$ for 
$\delta=0.9,\ 0.7,\ 0.5,\ 0.3,\ 0.1$ (from top to bottom). 
Here $\mu=1$ and $p=2.5$. The synchrotron spectrum 
with the same magnetic field strength is shown (dashed curve) for comparison.
The value of $\bar B^2_{SS}/\bar B^2_e$ is kept fixed for all $\delta$'s.
\label{fig:PrDelta} }
\efig
\bfig
\plotone{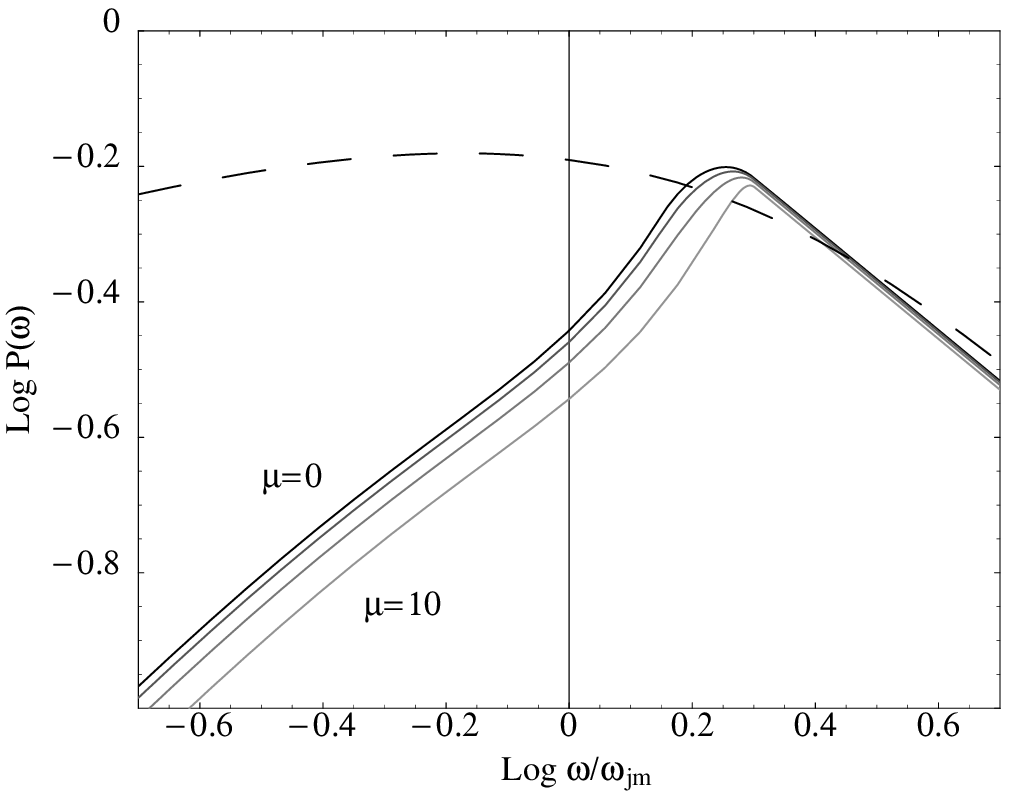}
\figcaption[PrEnsMu.eps]{
Spectral power $P(\omega)$ of radiation (log-log plot, arbitrary units) 
emitted by the power-law distributed electrons vs. $\omega/\omega_{jm}$ for 
$\mu=0,\ 1,\ 3,\ 10$ (from top to bottom).
Here $\delta=0.7$ and $p=2.5$. The synchrotron spectrum with the same magnetic 
field strength is shown (dashed curve) for comparison. 
The value of $\bar B^2_{SS}/\bar B^2_e$ is kept fixed for all $\mu$'s.
\label{fig:PrMu} }
\efig

Figure \ref{fig:PrDelta} represents the spectral power, $P(\omega)$, 
given by equations (\ref{Pr(w)1}), (\ref{Pens(w)}), as a function of 
$\omega/\omega_{cm}$, where $\omega_{cm}\equiv\omega_c(\gamma_{\rm min})$,
for various values of $\delta$. The dashed curve 
represents the synchrotron spectrum which corresponds to the same 
magnetic field strength. Note, the frequency is normalized by the 
synchrotron break frequency, $\omega_{cm}$ (not the jitter break frequency, 
$\omega_{jm}$), to emphasize that, for the fixed field magnitude, 
$\bar B_{SS}^2$, the jitter frequency increases with decreasing $\delta$, 
in accordance with equation (\ref{wcwj}). One can see that jitter 
radiation is well described by a broken power-law given by equation 
(\ref{asymptotics}). Notice also little change in the spectrum shape near the 
break frequency and the overall decrease of $P(\omega)$ as $\delta$ decreases.

Figure \ref{fig:PrMu} shows $P(\omega)$ vs. $\omega/\omega_{jm}$ for a 
few values of the magnetic field index, $\mu$. In general, the ratio 
$\bar B^2_{SS}/\bar B^2_e$ depends on $\mu$. To highlight the effect of the 
magnetic spectrum shape alone, we keep $\bar B^2_{SS}/\bar B^2_e$ fixed
for all $\mu$'s.  One can see from Figure \ref{fig:PrMu} that $\mu$ has 
little effect on the overall shape of the radiation spectrum. However,
as one goes from a flat ($\mu=0$) to a peaked ($\mu=10$) magnetic field 
distribution, the change in slope at the break frequency becomes more abrupt.

\subsection{Synchrotron radiation, \(\delta\gg1\) }

For completeness, we also consider the case of $\delta\gg1$.
In this case,  $\alpha/\Delta\theta\gg1$ while still $\lambda_B/\rho_L\ll1$ 
for energetic electrons. The spectral power emitted by a single electron is 
given as \cite[ \S 77 and \S 74]{LL}
\beq
\frac{dW}{dt\,d\omega}=\frac{e^2}{\sqrt{3}\pi c\gamma^2}
\left< \tilde\omega_c\;F\!\left(\frac{\omega}{\tilde\omega_c}\right)\right> ,
\label{Ps(w)}
\eeq
where $\tilde\omega_c=\frac{3}{2}\gamma^2e\tilde B_\bot/m_ec$,\
$\left<\dots\right>$ denotes the average over the electron's trajectory, 
and ``tilde'' denotes the local (instantaneous) quantity.
Here $F(\xi)=\xi\int_\xi^\infty K_{5/3}(\xi')\,d\xi'$ and $K_{5/3}(x)$ is
the modified Bessel function of $5/3$ order. As for the spectrum from a 
plasma, the observed radiation comes from regions with different field 
strength. Therefore, one has to replace $\tilde B_\bot$ in (\ref{Ps(w)}) 
with the ensemble average, $\bar B=\sqrt{\langle\tilde B^2_\bot\rangle}$, 
which rigorously yields the standard synchrotron spectrum.

\section{Jitter+Synchrotron model of GRB emission
\label{S:JS} }

We now use the model of magnetic fields discussed in \$ \ref{S:MMODEL}
to construct a GRB spectral model
All calculations are performed in the frame of the expanding 
shell. The transition to observer's frame is obvious.
The magnetic field in the shock shell was shown to be sub-divided into 
 small-scale and large-scale components. 
The small-scale field, for which $\delta\lesssim1$, yields jitter radiation.
Its spectral power is given by equations (\ref{Pr(w)1}), 
(\ref{Pens(w)}). The large-scale field, for which $\delta\gg1$, yields
synchrotron radiation. The relevant theory may be found in \citet{RL}.
The composite spectrum, thus, contains both jitter and synchrotron 
components and may be schematically represented as follows,
\beq
P_{J+S}(\omega)=P_J\left(\omega;\bar B_{SS},\delta,\mu\right)
+P_S\left(\omega;\bar B_{LS}\right) .
\eeq
This approximation is valid for the field distribution from \S \ref{S:MMODEL}
unless $\delta\sim1$, as discussed in the beginning of \S \ref{S:QT}.
It is convenient to normalize frequencies onto the jitter frequency, 
$\omega_{jm}$, which is independent of the magnetic field strength,
cf., equation (\ref{w-br}). The synchrotron-to-jitter frequency ratio
is then
\beq
\frac{\omega_{cm}}{\omega_{jm}}=\frac{\omega_c}{\omega_j}
\simeq\frac{3}{2}\frac{\bar B_{LS}}{\bar B_{SS}}\,\delta ,
\label{WcmWjm}
\eeq
as follows from equations (\ref{wcwj}), (\ref{Bls}).

\bfig
\plotone{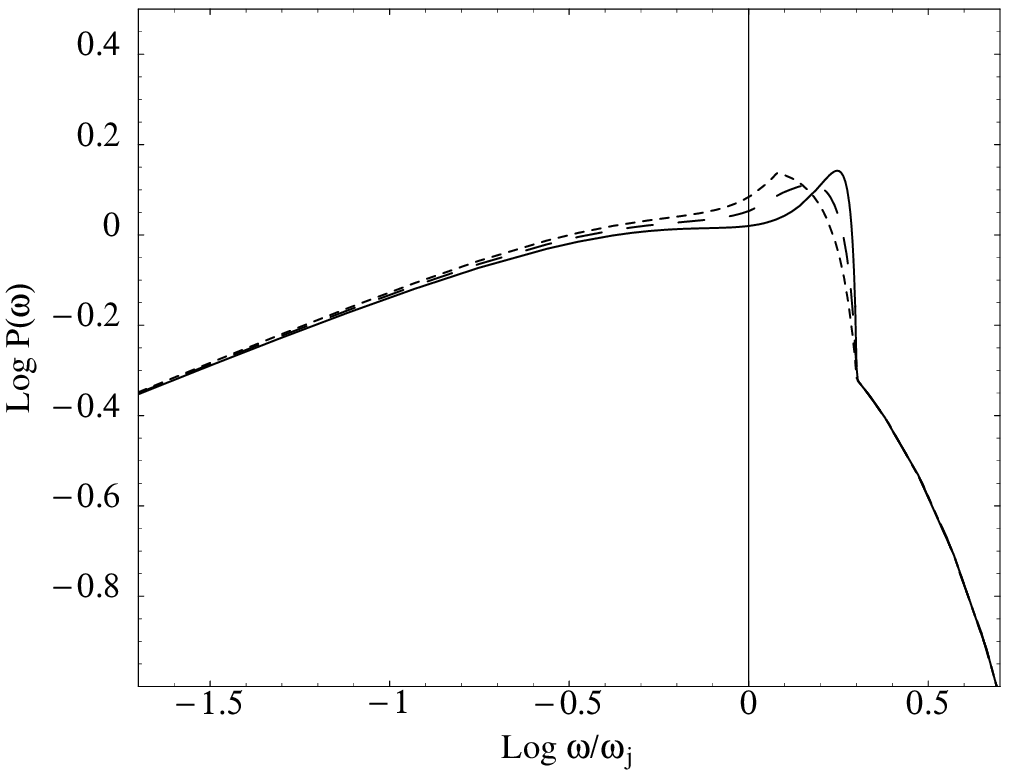}
\figcaption[PCompSingle.eps]{ The arbitrarily normalized, composite power 
spectra, $P_{J+S}(\omega)$, (log-log plot) emitted by a single electron
for several values of the magnetic field spectral index, $\mu=1$ (dotted),
$\mu=3$ (dashed), and $\mu=10$ (solid) for $\delta=0.6$ and 
$\bar B_0^2/\bar B_{SS}^2=3$.
\label{fig:CompSingle} }
\efig
\bfig
\plotone{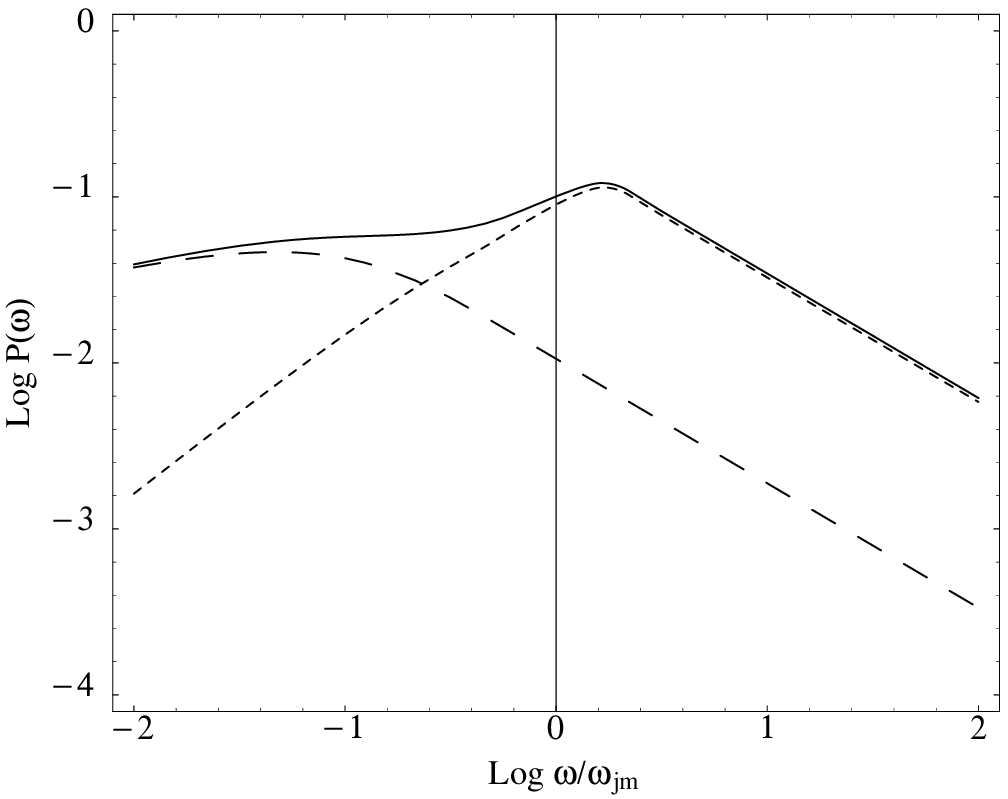}
\figcaption[PCompEnsem.eps]{ 
A typical composite power spectrum, $P_{J+S}(\omega)$, (log-log plot, 
arbitrary units) for the power-law distributed electrons with $p=2.5$ for 
$\mu=1$,\ $\delta=0.3$, and $\bar B_0^2/\bar B_{SS}^2=0$ is shown. 
The synchrotron (long-dashed curve) and jitter (short-dashed curve) 
sub-components are also shown.
\label{fig:exa} }
\efig

Figure \ref{fig:CompSingle} represents the spectrum emitted by single-speed 
electrons for the same values of $\mu$ as in Figure \ref{fig:PrSingle}.
One can clearly see a sharp feature on top of the broad synchrotron spectrum.
Integrating this spectrum over the power-law distribution of $\gamma$'s, 
we obtain the composite ``jitter+synchrotron'' (JS) model of GRBs.
A typical example is shown in Figure \ref{fig:exa}. In general, there are
two bumps: the sharp one is near the jitter frequency and the other, broad bump,
is associated with synchrotron emission. Depending on the ratio 
$\bar B_{LS}^2/\bar B_{SS}^2$, these bumps may overlap to produce either  
featureless broad or sharply peaked spectra, as well. 
The high-frequency tail always scales as 
$P_{J+S}(\omega)\propto\omega^{-(p-1)/2}$. The dependence of the
spectrum from $\delta$ is displayed in Figure \ref{fig:CompDelta}.
As an example, we take $\bar B_0^2/8\pi=\bar B_{SS}^2/8\pi$ and $\mu=10$
(for such $\mu$'s $\bar B_{SS}^2/8\pi\simeq\bar B_e^2/8\pi$, and, therefore
$\bar B_{LS}^2/8\pi\simeq\bar B_0^2/8\pi$). A sharp
jitter feature is easily seen in the spectrum. As $\delta$ decreases, 
the synchrotron bump moves towards lower frequencies and decreases in 
amplitude. The jitter peak decreases even faster and the spectral feature 
becomes less prominent. The other parameter, $\mu$, determines the
ratio $\bar B_{LS}^2/\bar B_{SS}^2$ but, besides this, its effect onto the 
spectrum is weak (cf., Figure \ref{fig:PrMu}) and is, therefore, not shown.

\bfig
\plotone{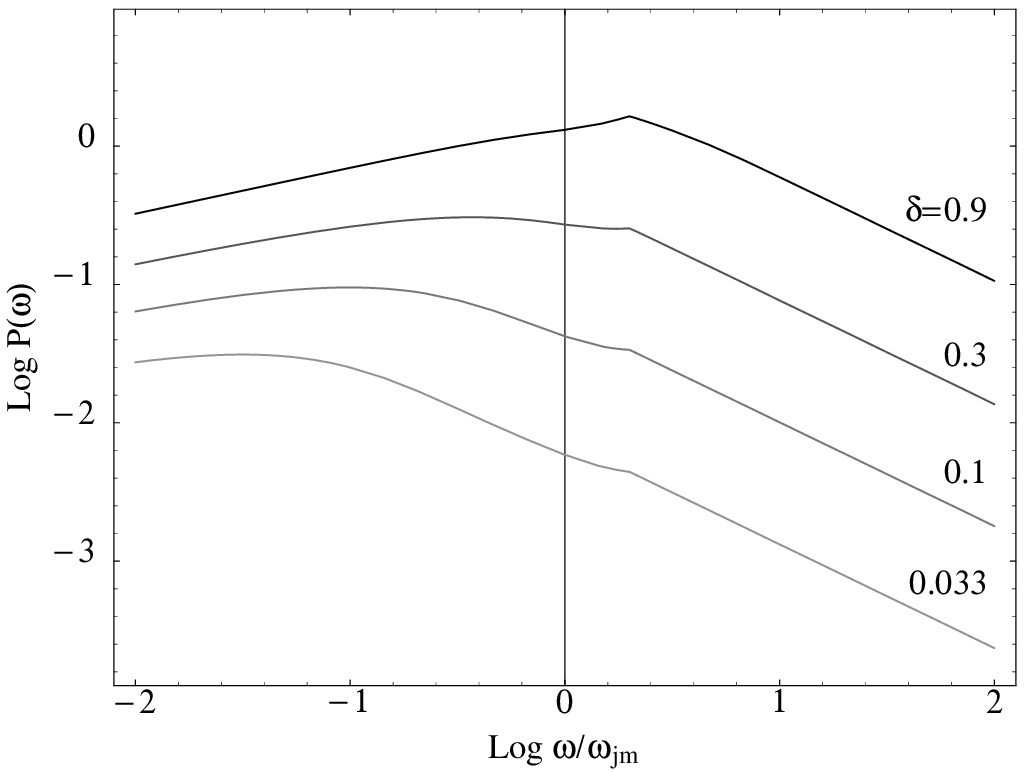}
\figcaption[PCompEns1Phet.eps]{ The composite power spectra, $P_{J+S}(\omega)$, 
(log-log plot, arbitrary units) for the power-law distributed electrons with 
$p=2.5$ for $\delta=0.9,\ 0.3,\ 0.1,\ 0.033$,\ $\bar B_0^2/\bar B_{SS}^2=1$ 
and $\mu=10$.
\label{fig:CompDelta} }
\efig

The spectrum of radiation depends on the relative magnitudes
of the large- and small-scale field components. Figure \ref{fig:CompEps} 
represents spectra for $\mu=10$ and various values of the ratio
$\bar B_0^2/\bar B_{SS}^2\simeq\bar B_{LS}^2/\bar B_{SS}^2$. 
Figure \ref{fig:CompEps}a shows the spectral power vs. frequency. 
For strong large-scale fields, the synchrotron spectral component dominates. 
As $\bar B_{LS}^2/\bar B_{SS}^2$ decreases, the synchrotron peak moves 
towards lower frequencies. The amplitude of the synchrotron peak decreases too.
The position and the amplitude of the jitter peak remain almost unchanged.
Thus, in the limit $\bar B_{LS}^2/\bar B_{SS}^2\to0$, the spectrum becomes
purely jitter. It is illustrative to depict the spectral flux, 
$F(\omega)\equiv P(\omega)/\omega$ [observationally, this quantity is
proportional to the photon spectral flux, $N(E)$, measured in units: 
photons~s$^{-1}$~cm$^{-2}$~keV$^{-1}$], as presented in 
Figure \ref{fig:CompEps}b. It is seen that the slope of the flux below
the jitter break continuously decreases as $\bar B_{LS}^2/\bar B_{SS}^2$
decreases. The synchrotron asymptotic slope $\propto\omega^{-2/3}$ is
shown for comparison. It is interesting that for 
$\bar B_{LS}^2/\bar B_{SS}^2\lesssim10^{-1}$ (the actual value depends on 
$\delta$; the larger $\delta$, the larger the ratio), there is a portion in
the spectrum which has the power-law index being less than $-2/3$ and
approaching zero in the limit of the vanishing large-scale field.
We discuss this property in \S \ref{S:D} in the context of the 
``line of death'' for synchrotron radiation in GRBs.

Finally, it is worthwhile to compare the peak spectral fluxes of jitter,
$F_{J,{\rm max}}\equiv F_J(\omega_{jm})$, and synchrotron, 
$F_{S,{\rm max}}\equiv F_S(\omega_{cm})$, radiation,
\beq
\frac{F_{J,{\rm max}}}{F_{S,{\rm max}}}=f(p,\mu)\,\delta^2,
\label{FJFS}
\eeq
where $f(p,\mu)$ is a function of two power-law indices, $p$ and $\mu$.
The above equation may be readily obtained from equations
(\ref{Pens(w)}), (\ref{Pr(w)1}), and, for instance, (\ref{Ps(w)}).
For $\mu\gg1$, the function $f$ depends on $\mu$ only weakly while, given
the spectrum, $p$ is found by fitting the large frequency slope.
Therefore, for a given spectrum with two sub-components, both significant
parameters, namely $\bar B_{LS}/\bar B_{SS}$ and $\delta$, are uniquely
found from equations (\ref{WcmWjm}) and (\ref{FJFS}).

\bfig
\plotone{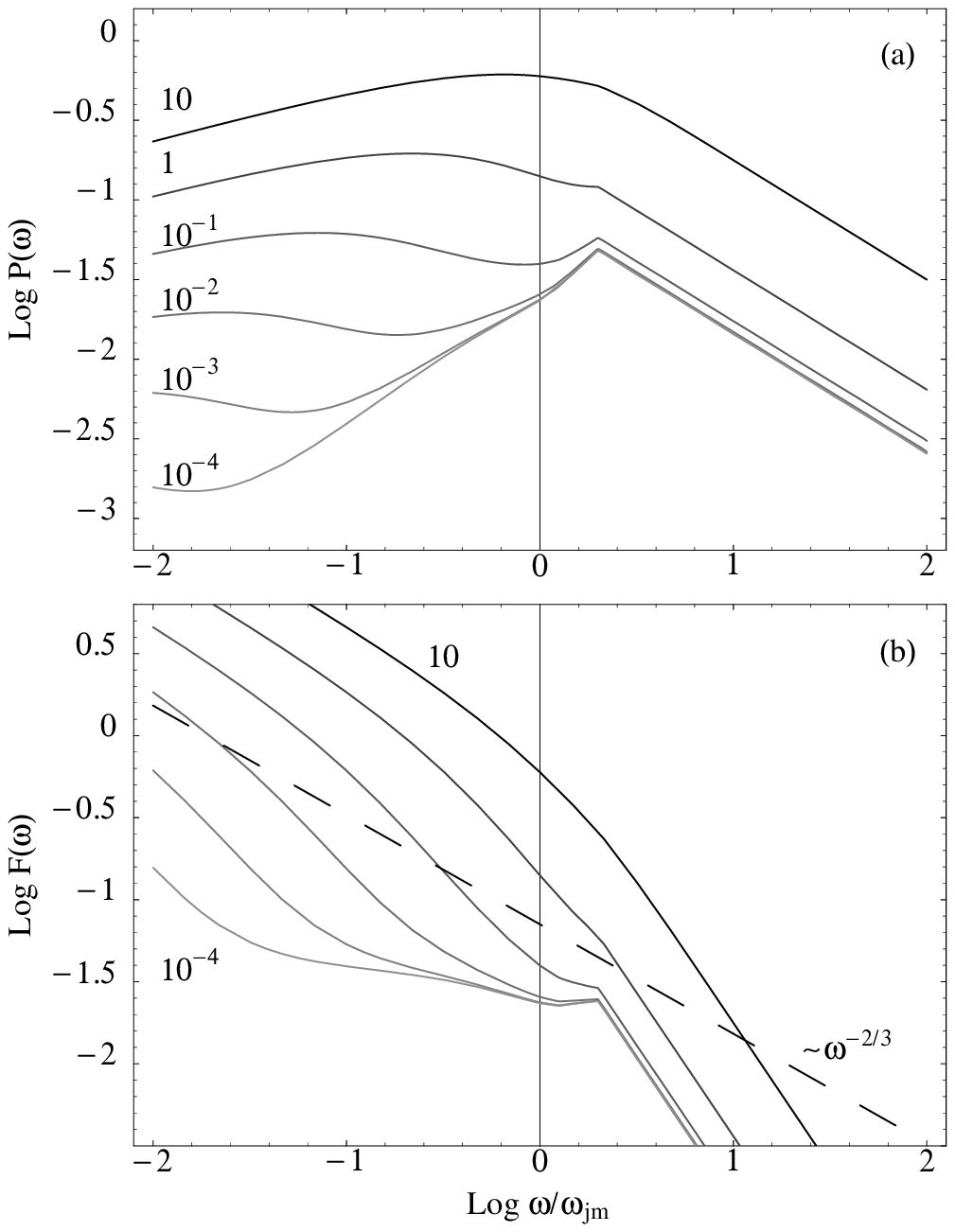}
\figcaption[PFCompEnsEpsB0.eps]{ 
The composite spectra (log-log plots, arbitrary 
units) for the power-law distributed electrons with $p=2.5$ for 
$\bar B_0^2/\bar B_{SS}^2=10-10^{-4}$,\ $\delta=0.2$,\ and $\mu=10$.
(a) --- Spectral power $P_{J+S}(\omega)$; 
(b) --- Spectral flux $F_{J+S}(\omega)\equiv P_{J+S}(\omega)/\omega$.
\label{fig:CompEps} }
\efig

\section{Comparison with observations
\label{S:D} }

In \S \ref{S:JS} we constructed a composite, jitter+synchrotron model 
of radiation emitted from internal shocks of GRBs, assuming that
the magnetic fields are produced in these shocks via the relativistic
two-stream instability \citep{ML99}. 
Several observational predictions can now be drawn.
We compare them with presently available observational data.
For future reference, we define the photon spectral index, $s$, as follows,
\beq
F(\omega)\propto\omega^s .
\eeq
Here we remind that $F(\omega)\equiv P(\omega)/\omega\propto N(E)$, where
$N(E)$ is the number of observed photons per unit time, per unit energy range,
per unit area.

We have shown that prompt GRB spectra consist of two spectral sub-components, 
namely synchrotron and jitter. The synchrotron component is, as usual, 
well approximated by the smoothly broken power-law, also referred to as 
the Band function \citep{Band93} or the GRB function. The jitter component
is better approximated by a sharply broken power-law with the hard and soft
photon indexes being equal to $s=-(p+1)/2$ and $s=0$ respectively.
The position of the jitter peak is independent of the magnetic field 
strength, in contrast to the synchrotron peak, but depends on the
particle (electron) density in the relativistic expanding shell, see
equation (\ref{w-br}).

\subsection{Low-frequency spectra and the ``Line of Death''}

The optically thin synchrotron model of GRBs makes a solid prediction. Namely,
the soft photon spectral index must be in the range $-3/2\le s\le-2/3$,
depending on the strength of the magnetic field if strong synchrotron cooling 
of the emitting electrons in the fireball is taken into account 
\citep{Katz94,SNP96}. Thus, the photon index in this model can 
newer be greater than $-2/3$, creating a testable ``line of death'' 
for the synchrotron shock model \citep{Katz94} .
There is, however, a growing observational evidence that many bursts
violate this prediction \citep{Crider97,Strohmayer98,Preece98,Frontera99}.
For instance, \citet{Preece98} have studied time-resolved spectra of the 
bursts collected by the LAD from the BATSE instrument. They found 23 
out of 137 bursts which violate the optically thin synchrotron model.
\citet{Frontera99} demonstrated that about $50\%$ of time-resolved 
spectra the bursts observed by the Wide Field Camera on board of the
{\it BeppoSAX} 
before May 1998 also violate this model. The proposed explanations, which 
include Compton up-scattering of low-energy photons \citep{Liang97}, 
synchrotron self-absorption \citep{Papathanassiou99}, and the influence of the
pair annihilation in the fireball photosphere \citep{EL99}, though possible, 
seem rather {\it ad hoc} and suffer from drawbacks. The Compton 
up-scattering model strictly requires a single up-scattering event 
per photon which requires the column density to self-adjust to a few g/cm$^2$.
The self-absorption model results in large optical depths and, thus, weak 
emitted flux and very low radiation efficiency. The photospheric model requires
an extremely low baryonic load. For more discussion see \citet{Mes99}.

\bfig
\plotone{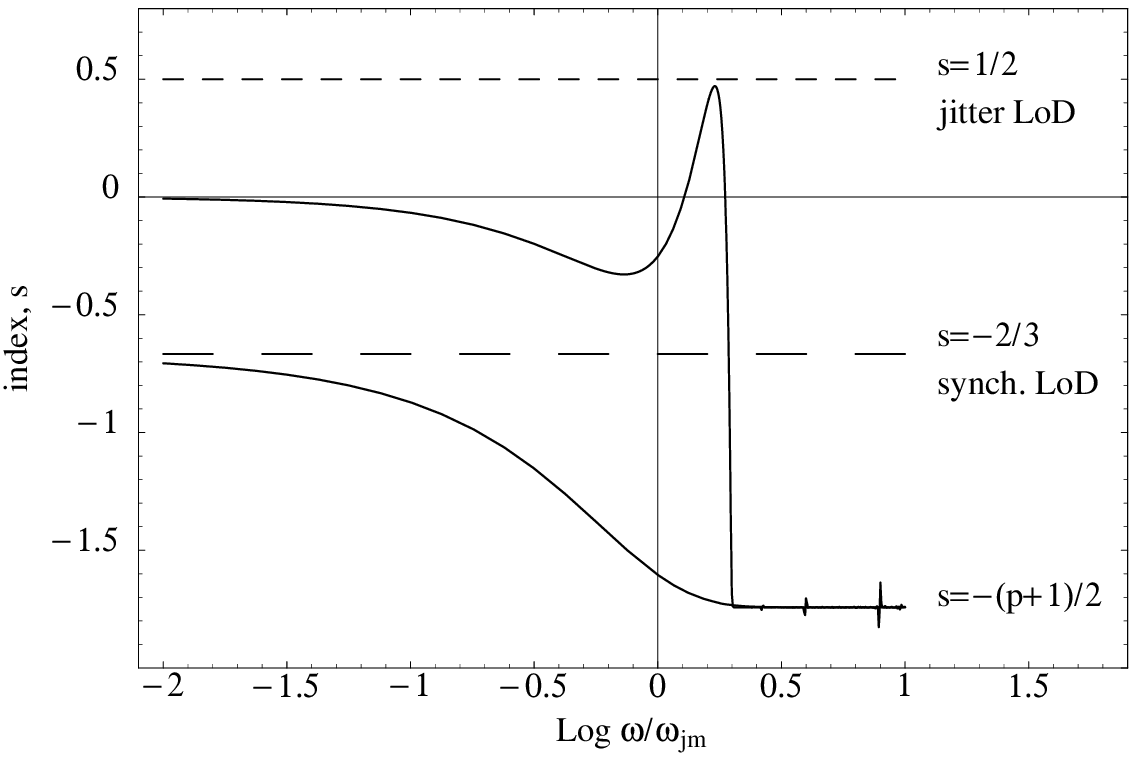}
\figcaption[IndexS.eps]{ The power-law index, $s$, vs. 
$\log(\omega/\omega_{jm})$ for synchrotron and jitter radiation with $\mu=10$,
$\delta=0.2$, $\bar B_{LS}^2/\bar B_{SS}^2=1$.
The synchrotron and jitter ``lines of death'' (LoD) are shown.
\label{fig:index} }
\efig

The JS model provides a natural resolution of this puzzle. 
Figure \ref{fig:index} represents the photon index as a function of a 
logarithm of frequency for the synchrotron and jitter cases%
\footnote{Little spikes on the lower curve are numerical artifacts.} 
When jitter radiation dominates, the low-energy photon index tends 
towards its asymptotic value of $s=0$, thus shifting the
``line of death'' to harder spectra. In principle, the indexes as large as
$s=1/2$ are allowed by our model. Positive $s$, however, are quite unlikely.
These predictions are in good agreement with observational data. 
Namely, only two bursts out of those studied by \citet{Preece98} are
above the $s=0$ line by more than $1\sigma$ and all bursts studied by
\citet{Frontera99} except the
peculiar one, GRB~970111, fall below this line. At last, these three
bursts are still consistent with $s=1/2$ within statistical uncertainties.

\subsection{Sharply broken power-law spectra of GRBs}

Most of the observed bursts are well-fitted by the GRB function \citep{Band93}. 
However, there are bursts with a sharp curvature of the spectrum at the break 
energy. For such bursts a broken power-law model generates better $\chi^2$
fits than the GRB function (see, e.g., \citealp{Preece98}). In terms of the 
JS model, there are two cases in which such spectra occur. They can be 
observationally distinguished by the value of the soft (i.e., low-energy) 
photon index. First, it is the case of purely jitter radiation:
$\bar B_{LS}^2/\bar B_{SS}^2\to0$ and there is no synchrotron peak 
(or this peak is too weak and is at low energies, outside the detector range), 
see Figure \ref{fig:CompEps}. These spectra are flat, $s\sim0$, at low 
energies. Second, for large $\delta\sim1$ and the equipartition between the 
large- and small-scale fields, $\bar B_{LS}^2/\bar B_{SS}^2\sim1$, similar 
spectral shapes are also obtained, see Figure \ref{fig:CompDelta}. 
In this case, the soft photon index is $s\sim-2/3$, or it is in the
range $-3/2\le s\le-2/3$ if synchrotron cooling of the electrons is taken
into account \citep{SNP96}. Thus, we expect fewer broken power-law bursts 
which have the soft photon indexes in the range $-2/3\lesssim s\lesssim0$.
Figure 2 of \citet{Preece98} indeed reveals a gap between the power-law 
bursts with $s\sim0$ and $s\lesssim-2/3$. However, a larger number of such 
bursts is required to draw a statistically significant conclusion.
We also should point out that if no ordered field is present in the ejecta,
then $\bar B_{LS}\sim\bar B_{SS}$ is equivalent to 
$\bar B_p^2/8\pi\sim\bar B_e^2/8\pi$. This, in turn, naturally implies
strong energy coupling and rough equipartition between the protons and 
the electrons, as discussed in \S \ref{S:MEP}.

\subsection{Two-component spectral structure of GRB emission}

The analysis done by \citet{Pendl94} of the bursts from the first BATSE
catalog collected by the Large Area Detector (LAD) demonstrates that there is 
a large number of bursts which have the high-energy photon indexes 
($50-300$~keV range) being much larger than the low-energy ones 
($20-100$~keV range). No such behavior is observed at other energy ranges. 
Thus, this result may indicate the presence of more than one spectral 
component in the energy range $40-100$~keV. (The absence of similar 
behavior at higher energies seems to rule out the inverse-Compton origin 
of a second spectral component.) On the other hand, such spectra are 
completely consistent with the JS model. 
They generally correspond to the well-separated synchrotron and jitter peaks 
and are characterized by  small $\delta$'s, $\delta\lesssim0.1$, and weak 
large-scale fields, $\bar B_{LS}^2/\bar B_{SS}^2\lesssim0.01$, as is 
clearly seen from Figure \ref{fig:CompEps}.

\begin{figure*}[t]
\epsscale{1.4}\plotone{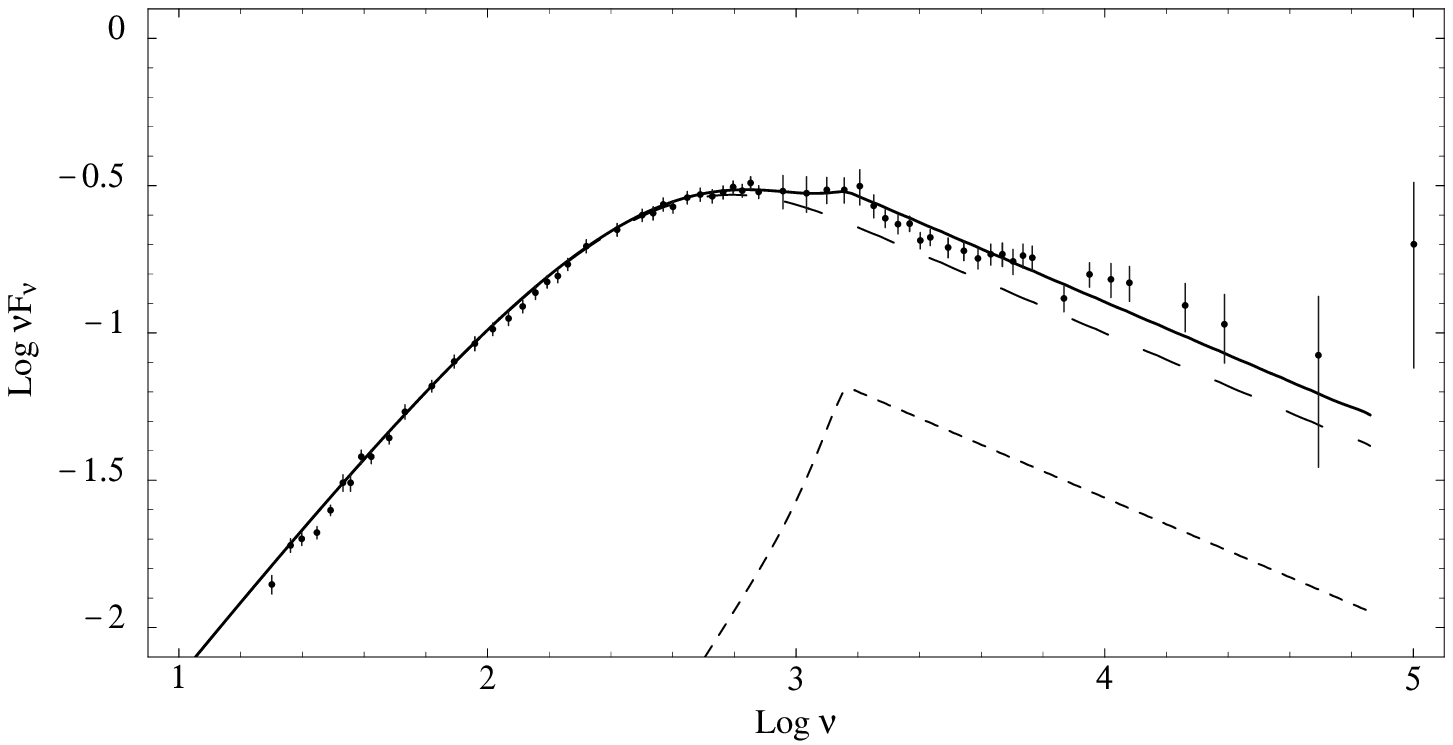}
\caption[FitGRB910503.eps]{ A visual fit of a spectral shape for GRB~910503.
The inferred values are: $\bar B_0^2/B_{SS}^2=7,\ \delta=0.07,\ \mu=10,\ p=3.9$.
The synchrotron (long-dashed curve) and jitter (short-dashed curve) spectral
sub-components are also shown.
\label{fig:fit} }
\end{figure*}

\subsection{``Lines'' in GRB spectra}

Whether emission and/or absorption features, often referred to as lines, 
in GRB spectra are real remains
an open question for a long time \citetext{see, e.g., \citealp{Briggs99} and
\citealp{Ryde99} for discussion}. These features have been observed, for
instance, by the KONUS experiment, {\it Venera} mission \citep{Mazets81}
and by {\it Ginga} \citep{Murakami88}. At that time, it was widely believed 
that these features are cyclotron lines due to a highly magnetized progenitor.
The BATSE Spectroscopy Detectors (SD) refined these results. 
\citet{Palmer94} studied almost two hundred bursts detected by SDs. 
No convincing line features were found. In a more recent analysis by
\citet{Briggs99} of more than a hundred bursts observed by SDs before
May 31, 1996, about 10 highly significant line features have been found.
These are low-energy ($\sim50$~keV) emission features. Clearly, this 
controversy requires further studies. However, it seems likely that 
if lines exist, they are rare.

Not too surprisingly, the two-component JS model is able to produce 
spectra with emission line-like features. These are not lines in a strict 
sense because at energies higher than the ``line'' peak, the spectrum has
no curvature and falls down as a power-law with $s=-(p+1)/2$.
Few such spectra are seen in Figure \ref{fig:CompDelta}. These spectra
are obtained (i) for rather low values of $\delta$ ($\delta\lesssim0.3$),
otherwise the synchrotron and jitter peaks overlap, and (ii) for a narrow
range of $\bar B_{LS}^2/\bar B_{SS}^2$ 
($0.1\lesssim\bar B_{LS}^2/\bar B_{SS}^2\lesssim1$), otherwise either the 
spectral components are too well separated if the value of this ratio is small, 
or the jitter peak is too week and unobservable if the ratio is large.
We thus conclude that the spectra with emission features are not very
common. They require some parameter tuning and, therefore, must be rare.
Moreover, inhomogeneities and highly variable conditions in the GRB 
shocks may smear out weak spectral features completely.

At last, we present an illustrative example. Figure \ref{fig:fit} shows 
the spectrum of GRB~910503 obtained using all capability of {\it CGRO}'s
four experiments \citep{Schaefer98}. Here the energy flux 
$\nu F_\nu\propto E^2N(E)\propto\omega P(\omega)$ in units of
(photons cm$^{-2}$ s$^{-1}$ keV$^{-1}$)$\times(E/100$~keV)$^2$ is plotted
vs. frequency in units of keV. A sharp second peak at $\sim2$~MeV is clearly 
seen. The solid curve in the figure represents a visual (i.e., not $\chi^2$) 
spectral shape fit using the JS model. Here we fit the separation of the 
two peaks and their relative amplitudes; we didn't fit the absolute flux
and position of the synchrotron spectral break which depend on the
energetics of the fireball. Note that the intensity of a second, 
high-frequency (jitter) component is low. The values inferred from 
this fit are the following:
$\bar B_{LS}^2/\bar B_{SS}^2=7$,\ $\delta=0.07$,\ $p=3.9$, and we took
$\mu=10$. We calculate the efficiency of the magnetic field generation 
in the shock from equation (\ref{delta}) assuming $\phi=1/4$ as a 
typical value. We obtain $\eta_e\simeq0.08$. The total magnetic field 
energy is calculated as
$\epsilon_B=\epsilon_{Be}(1+\bar B_{LS}^2/\bar B_{SS}^2)$, where 
$\epsilon_{Be}$ is found from equation (\ref{Be}) and we used
$\bar B_{SS}\simeq\bar B_e$ for large $\mu$. We obtain
$\epsilon_B\simeq4\times10^{-4}$, which is in agreement with the
conclusion drawn by \citet{Dermer99} from a completely different 
analysis that the magnetic field in this burst
is well below the equipartition, $\epsilon_B\lesssim10^{-2}$.
Their argument goes from the fact that the low-energy spectrum of 
this burst is consistent with $F(\omega)\propto\omega^{-2/3}$.
Evidently, synchrotron losses are small and 
the emitting electrons do not form a cooled distribution, which 
otherwise would result in $F(\omega)\propto\omega^{-3/2}$. This is
possible only for very small values of $\epsilon_B$.

\section{Conclusion
\label{S:C} }

In this paper we have shown that radiation produced by relativistic 
electrons in magnetic fields may differ quite substantially from synchrotron.
This radiation, referred here to as jitter radiation, is produced 
in the magnetic fields which are highly inhomogeneous on very small spatial
scales. Such fields are likely to be present in GRB shocks. We developed
a quantitative theory of jitter radiation. Jitter radiation has a different 
spectrum and its peak frequency is independent of the field strength. However, 
the total (i.e., frequency integrated) emitted power of jitter radiation 
depends on the field strength and is exactly identical to that of synchrotron 
radiation. We also constructed a composite, two-component jitter+synchrotron 
spectral model of the prompt GRB emission. Predictions of this model seem to 
be in excellent agreement with presently available data and, likely, resolve 
some puzzling spectral properties of the prompt $\gamma$-ray emission. 
All this, we think, strongly supports
that (i) the proposed jitter radiation mechanism operates in astrophysical 
objects and (ii) the magnetic field is generated in shocks by the two-stream 
instability. We emphasize that a reliable identification/detection of the
jitter spectral features will provide a direct evidence that the magnetic 
field in GRBs is due to the two-stream  instability, since we presently unaware 
of any other mechanism which is capable of producing the required small-scale,
large-amplitude fields.
In general, the detection of both spectral components in GRB 
spectra would be a powerful and presize tool to investigate the properties of 
cosmological fireballs. 

It is important to emphasize that the phenomenon of jitter radiation is
intrinsic not only to internal shocks. Similar conditions (i.e., strong, 
small-scale fields) are expected to occur in external shocks which produce 
delayed afterglows, as well as in more conventional supernova shocks
and relativistic jets. More speculatively, the above mechanism could allow
to study the magnetic and electric fields in reconnection regions
(remember, reconnection occurs on the electron skin depth scales)
and small-scale structure of magnetic turbulence and cascade 
(e.g., in the interstellar medium). 

The theory presented in this paper is only a ``first step''. It is incomplete
in a sence that it does not include the effects of both ordered and
large-scale random magnetic fields in a self-consistent way and, hence, 
does not smoothly interpolate between the two extremes of synchrotron 
and jitter mechanisms. The synchrotron self-absorption has been omitted.
The model also does not consider some related effects. For instance,
it is likely that random electric fields (i.e., strong Langmuir turbulence)
are present in the cosmological collisionless shocks. These fields will
definitely affects the radiation spectrum via a similar mechanism. 
It is also unclear now whether particle's motion is ergodic in random fields
(i.e., whether a particle ``samples'' strong and weak fields statistically 
homogeneously), and what could be the effect of nonergodicity on the observed 
spectrum. All these issues will be addressed in future publications.

\acknowledgements

The author is grateful to Ramesh Narayan and Norm Murray for their interest
in this work, various insightful comments, and useful discussions and to
George Rybicki and Avi Loeb for discussions. This work has been supported 
by NSF grant PHY~9507695.

\end{document}